\documentclass[notoc]{JHEP} 

\usepackage{epsfig}
\usepackage{latexsym}

\newcommand\fverb{\setbox\pippobox=\hbox\bgroup\verb}
\newcommand\fverbdo{\egroup\medskip\noindent%
			\fbox{\unhbox\pippobox}\ }
\newcommand\fverbit{\egroup\item[\fbox{\unhbox\pippobox}]}
\newbox\pippobox

\def\ra{\rightarrow}
\def\lsim{\raise0.3ex\hbox{$\;<$\kern-0.75em\raise-1.1ex\hbox{$\sim\;$}}}
\def\gsim{\raise0.3ex\hbox{$\;>$\kern-0.75em\raise-1.1ex\hbox{$\sim\;$}}}
\def\beq{\begin{equation}}
\def\eeq{\end{equation}}
\def\nt{\hbox{$\nu_\tau$ }}
\def\VEV#1{\left\langle #1\right\rangle}
\def\eq#1{{eq.~(\ref{#1})}}
\def\fig#1{{Fig.~\ref{#1}}}
\def\apj#1#2#3{{\it Astrophys. J. }{\bf #1} (#2) #3}
\def\ppnp#1#2#3{{\it Prog. Part. Nucl. Phys.} {\bf #1} (#2) #3}

\title{Lepton flavour violation in a left-right symmetric model}

\author{Sergio Pastor\\
SISSA/ISAS, via Beirut 2-4, 34013 Trieste, Italy\\
E-mail: \email{pastor@sissa.it}}
\author{Saurabh D. Rindani\\
Theory Group,
Physical Research Laboratory\\
Navrangpura, Ahmedabad 380 009, India\\
E-mail: \email{saurabh@prl.ernet.in}}
\author{Jos\'e W. F. Valle\\
Instituto de F\'{\i}sica Corpuscular - C.S.I.C.\\
Departament de F\'{\i}sica Te\`orica, Universitat de Val\`encia\\
46100 Burjassot, Val\`encia, Spain\\
E-mail: \email{valle@flamenco.ific.uv.es}}

\preprint{hep-ph/9705394}

\abstract{
We consider in this paper a Left-Right symmetric gauge model in which
a global lepton-number-like symmetry is introduced and broken
spontaneously at a scale $\omega$ that could be as low as $10^4$ GeV
or so. The corresponding physical Nambu-Goldstone boson, which we call
majoron and denote $J$, can have tree-level flavour-violating couplings
to the charged fermions, leading to sizeable majoron-emitting
lepton-flavour-violating weak decays.  We consider explicitly a
leptonic variant of the model and show that the branching ratios for
$\mu \to e+J$, $\tau \to e + J$ and $\tau \to \mu + J$ decays can be
large enough to fall within the sensitivities of future $\mu$ and
$\tau$ factories. On the other hand the left-right gauge symmetry
breaking scale may be as low as few TeV.
}

\keywords{Spontaneous Symmetry Breaking, Global Symmetries, Beyond
Standard Model, Rare Decays}

\begin{document} 

\section{Introduction}

Apart from offering a possibility of understanding parity violation on
the same footing as gauge symmetry breaking, left-right symmetric
extensions of the standard electroweak theory naturally incorporate
small neutrino masses \cite{LR1,LR2}.  Unfortunately, for
phenomenologically interesting values of the scale at which the
$SU(2)_R$ symmetry gets broken ($\lsim $ 10 TeV) one expects the
neutrino masses to lie close to their present laboratory limits,
unless the neutrino Yukawa couplings are suppressed.

It is well known that, if stable, neutrinos would contribute too much
to the energy density of the universe if their mass exceeds 60 eV or
so \cite{RHOCRIT}. Thus heavy neutrinos can be consistent with
cosmology only if there are new neutrino decay and/or annihilation
channels in addition to those induced by the standard model
interactions.

The most attractive possibility to provide such new interactions is
realized in models where the neutrino mass arises from the spontaneous
violation of a global $B-L$ symmetry \cite{CMP}.  In this case there
are invisible neutrino decays by majoron emission which can reconcile
the large neutrino masses with cosmological density arguments
\cite{V,CON,fae}. Moreover, the new majoron decay and/or annihilation
channels may also make these model consistent with cosmological
Big-Bang nucleosynthesis \cite{DPRV,bbnothers}, while avoiding
conflict with astrophysics \cite{Raffelt0}.

Unfortunately, since $B-L$ is a gauge symmetry, its spontaneous
breaking in the left-right symmetric framework does not imply the
existence of a majoron and, as a result, the possibility of fast
neutrino decay and/or annihilations is absent. Consequently, neutrino
masses and the scale of left-right symmetry breaking are strongly
restricted.

In order to cope with this problem a variant of the left-right
symmetric model with an additional spontaneously broken U(1) global
symmetry was proposed in ref. \cite{previous}.  In order to implement
such symmetry one adds new $SU (2)_L$ singlet leptons on which it acts
nontrivially. These isosinglet leptons mix with the ordinary neutrinos
and are involved in generating their mass. As a result, though the
U(1) global symmetry differs from the usual $B-L$ symmetry, it plays
an important role in generating the neutrino masses and decays.
Extending the majoron concept to cover this situation, we still call
majoron the Nambu-Goldstone boson that follows from the breaking of
the U(1) global symmetry. It has been shown in ref.
\cite{previous} that majoron-emitting neutrino decays e.g. \nt $\ra
\nu + J$ can easily reconcile the large neutrino masses and the low
left-right symmetry breaking scales with the cosmological
restrictions.

In this paper we extend this class of models also in the charged
fermion sector by adding electrically charged $SU (2)_L$ singlets.
These have been widely discussed in the case of baryon-number and
lepton-number-carrying SU(2) singlets, so-called lepto-quarks, both in
the context of superstring inspired models \cite{lqsst} or simply
added to the electroweak theory directly \cite{lqphen,lqphen2}. Models
with vector-like isosinglet quarks and leptons have also been invoked
in the context of models with universal seesaw mechanism to explain
the fermion masses \cite{univseesaw}.  On the other hand their
phenomenological implications as well as the prospects for searching
at high energy colliders have also been the subject of dedicated
experimental study groups.  In the present paper we consider, for
simplicity the case of leptons in detail. This extension can have
interesting phenomenological implications.  We pay special attention
to the flavour non--diagonal majoron couplings to usual charged
leptons, electron, muon and tau. These would lead to the existence of
majoron-emitting lepton-flavour-violating weak decays,
\beq
\mu \to e+J    \qquad
\tau \to e + J \qquad
\tau \to \mu + J
\label{single}
\eeq
(here $J$ denotes the majoron which follows from the spontaneous 
violation of the U(1) symmetry). Such decays have already been 
considered in other contexts, see e.g. ref. \cite{NPBTAU}. They 
would lead to bumps in the final lepton energy spectrum, at half 
of the parent lepton mass. These decays have been searched for 
experimentally \cite{mu,tau} but the limits on their possible 
existence are still rather poor, especially for the case of taus. 

Restricting ourselves to the case where the global symmetry 
breaking scale is the largest, we derive analytically by two
alternative methods, the direct one and the Noether current method
of ref. \cite{774}, the general structure of the couplings of the
charged leptons to the majoron, focusing on the off-diagonal couplings
that lead to the decays above.  We also consider the more
general case where the hierarchy of scales is relaxed.  Our numerical
calculations show that these decays may have branching ratios which
fall within the sensitivities of future $\mu$ and $\tau$ factories
\cite{CLEO97,9307252}.

\section{The Model}

We consider a model based on the gauge group
$$
G_{LR} \equiv SU (2)_L \otimes SU (2)_R \otimes U (1)_{B - L}
$$
in which an extra $U(1)_G$ global symmetry is postulated.
In addition to the conventional quarks and leptons,
there is a neutral gauge singlet fermion 
\footnote{Although the number of such singlets is arbitrary,
since they do not carry any anomaly, we add just one such
lepton in each generation, while keeping the quark sector as
the standard one. }.
These extra electrically neutral leptons might arise in superstring
models \cite{SST1}. They have also been discussed
in an early paper of Wyler and Wolfenstein \cite{WYLER}.
We do not use the more conventional triplet higgs scalars,
which are absent in many of these string models. Instead
we will substitute them by the doublets $\chi_{L}$ and $\chi_{R}$.

\TABLE[t]{
\begin{tabular}{|c|c|c|c|c|}
\hline
 & $ SU (2)_{L} \; \otimes$ & $SU (2)_{R} \; \otimes$ 
& $U(1)_{B - L} \; \otimes$ & $U(1)_G$\\
\hline \hline
$Q_{Li}$ & 2 & 1 & 1/3 & 0\\[0.1cm]
$Q_{Ri}$ & 1 & 2 & 1/3 & 0\\[0.1cm]
\hline
$\psi_{Li}$ & 2 & 1 & $-1$ & 0\\[0.1cm]
$\psi_{Ri}$ & 1 & 2 & $-1$ & 0\\[0.1cm]
$E_{Li}$ & 1 & 1 & $-2$ & $-1$ \\[0.1cm]
$E_{Ri}$ & 1 & 1 & $-2$ & 1 \\[0.1cm]
$S_{Li}$ & 1 & 1 & 0 & 1\\[0.1cm]
\hline
$\phi $& 2 & 2 & 0 & 0\\[0.1cm]
$\chi_{L}$ & 2 & 1 & $-1$ & 1\\[0.1cm]
$\chi_{R}$ & 1 & 2 & $-1$ & $-1$\\[0.1cm]
$\sigma$ & 1 & 1 & 0 & 2\\[0.1cm]
\hline
\end{tabular}
\caption{$SU (2)_L \otimes SU (2)_R \otimes U (1)_{B-L}
\otimes U(1)_G$ assignments of the quarks, leptons and higgs
scalars.}
\label{table1}}
In addition to this fermion content which was discussed in the earlier
model \cite{previous}, we introduce one $SU(2)_L \otimes SU(2)_R $
singlet charge-$-1$ lepton $E_i$ for each generation. These carry
$B-L$ of $-2$.  The left-handed and right-handed components of $E_i$
carry the global $U(1)_G$ charges of $-1$ and $+1$ respectively.  The
matter and Higgs boson representation contents are specified in Table
\ref{table1}.  With these assignments, the most general Yukawa interactions
allowed by the $G_{LR} \otimes U(1)_G$ symmetry are given as
$$
- {\cal L}_Y  = g_1 \bar{Q}_L \phi Q_R + g_2 \bar{Q}_L \tilde{\phi} Q_R
+ g_3 \bar{\psi}_L \, \phi \,  \psi_R +
g_4 \bar{\psi}_L \tilde{\phi} \psi_{R}
 + g_5 [\bar{\psi}_{L} \chi_L S^c_R + \bar{\psi}_R \chi_R S_L ] 
$$
\beq
+g_6 \bar{S}_L S^c_R \sigma
+ ig_7 \left( \bar{\psi}_L\tau_2 \chi_L^* E_R -
\bar{E}_L\chi_R^T\tau_2\psi_R
\right) + g_8 \bar{E}_LE_R\sigma^* 
+  h.c.
\eeq
where $g_i$ are matrices in generation space and
$\tilde{\phi} = \tau_2 \phi^* \tau_2 $ denotes the
conjugate of $\phi$. This Lagrangian is invariant
under parity operation $Q_L \leftrightarrow Q_R$, $\psi_L \leftrightarrow 
\psi_R$, $E_L \leftrightarrow E_R$,
$S_L \leftrightarrow S^c_R$, $\phi \leftrightarrow \phi^\dagger$ 
and $\chi_L \leftrightarrow \chi_R$.

The symmetry breaking pattern is specified by the
following scalar boson vacuum expectation values (VEVs), assumed real:
\beq
\VEV{\phi} =  \left( \begin{array}{cc}
k & 0 \\
0 & k'
\end{array} \right) \:;
\VEV{\chi_L} = \left( \matrix{v_L \cr 0} \right) \:;
\VEV{\chi_R} = \left( \matrix{v_R \cr 0} \right) \:;
\VEV{\tilde{\phi} } = \left(
\begin{array}{cc}
k' & 0 \\
0 & k
\end{array} \right) \:;
\VEV{\sigma } = \omega
\eeq
The spontaneous violation of the global $U(1)_G$ symmetry generates
a physical majoron whose exact profile 
is specified as 
\beq
\label{profi}
\left( \omega^2 + \frac{v_L^2 v^2}{V^2 r} \right)
 ^{- \frac{1}{2} } \left\{ \omega
 \sigma_I  + \frac{1}{V^2 v_R^2 r} \Bigl [v_R^2 v^2 (v_L  \chi_L^I)
 -v_L^2 v^2 (v_R \chi_R^I) - v_L^2 v_R^2 (k  \phi_2  - k' \phi_4 )
\Bigr ] \right\}
\eeq
where  $\sigma_I$, $\chi_L^I$, $\chi_R^I$, $\phi_2$ and $\phi_4$
denote the imaginary parts of the neutral fields
in $\sigma$, $\chi_L$, $\chi_R$ and the bidoublet $\phi$.
Here we have also defined the VEVs as
$v^2 \equiv k^2 + k'^2$ and $V^2 \equiv v^2 + v_L^2$, 
and 
\beq
r = 1 + \frac{v^2 v_L^2}{V^2 v_R^2}.
\eeq

Clearly, as it must, the majoron is orthogonal to the Goldstone bosons
eaten-up by the $Z$ and the new heavier neutral gauge boson $Z'$
present in the model. The latter acquires mass at the larger scale
$v_R$.

In the limit $V^2 \ll \omega^2 \ll v_R^2$, the majoron becomes
\beq
\label{profilargevr}
 J   = \left ( \omega^2 + \frac{v_L^2 v^2}{V^2} \right)^{- \frac{1}{2}} 
\left\{ \omega
 \sigma_I  + \frac{v_L v}{V^2} \left [ v  \chi_L^I
 - \frac{v_L}{v} (k  \phi_2  - k' \phi_4 )\right ] \right\}
\eeq

Note that in this limit, the majoron has no component along the imaginary
part of $\chi_R$ despite the fact that $\chi_R$ is
nontrivial under the global symmetry. This was the limit considered in 
\cite{previous}.

Here, in order to determine analytically the light charged-lepton
masses and majoron couplings, we will restrict ourselves to the case
where $\omega$ is the largest scale. In the
limit $V^2  \ll v_R^2 \ll \omega^2$, and
$r \rightarrow 1$, the  majoron field becomes
\beq
\label{profilargew}
 J   = \left( \omega^2 + \frac{v_L^2 v^2}{V^2} \right)
^{-\frac{1}{2}}    \left \{ \omega
 \sigma_I  + \frac{1}{V^2} \Bigl [ v^2 (v_L \chi_L^I)- 
\frac{v_L^2 v^2}{v_R^2} (v_R \chi_R^I)
 - v_L^2 (k  \phi_2  - k' \phi_4 )\Bigr ] \right \}
\eeq

The various scales appearing in \eq{profilargew} are not
arbitrary. First of all, note that the minimization
of the scalar potential dictates the consistency relation
\cite{LR2}
\beq
\label{vevseesaw}
\frac{v_R}{v} \sim \lambda \frac{\omega }{v_L}
\eeq
For $\lambda \sim 1$ the singlet VEV is necessarily
larger than $v_L$ i.e. $ v_L \ll \omega$ and, as a result,
the majoron is mostly singlet and the invisible decay  of
the $Z$ to the majoron is enormously suppressed, unlike
in the purely doublet or triplet majoron schemes.

On the other hand, in order that majoron emission
does not overcontribute to stellar energy loss 
one needs to require \cite{DEAR} in the general case
\beq
\label{red0}
\frac{g_{eeJ}}{m_e} \lsim 10^{-9}~ \mbox{GeV}^{-1}\;,
\eeq
where $g_{eeJ}$ is the majoron diagonal coupling to electrons.
In the limit $V^2  \ll v_R^2 \ll \omega^2$, and
$r \rightarrow 1$, one needs to require (see \eq{diagandoff})
\beq
\label{red}
\frac{v_L^2}{\omega V^2} \lsim 10^{-9}~ \mbox{GeV}^{-1}\;,
\eeq
One sees that \eq{vevseesaw} and \eq{red}
allow for the existence of right-handed weak interactions at
accessible levels, provided $v_L$ is sufficiently small.

The diagonal couplings of the majoron, for example to 
electrons, may also  have important implications
in cosmology and astrophysics \cite{fae}.

\section{Charged Lepton Masses}

Once all gauge and global symmetries get broken,
mass terms are generated for the charged     
leptons, which are given by
\beq
\label{yukawas}
- {\cal L}_{mass}  = 
  \bar{l}_L (g_3 k' + g_4 k)l_R 
-  \bar{l}_L g_7 v_L E_R 
 -  \bar{E}_L g_7 v_R l_R 
+  \bar{E}_L g_8 \omega E_R  + h.c.
\eeq
It may be written in block form as
\beq
\label{MAT}
- {\cal L}_{mass}  = 
\left(
\begin{array}{cc}
\bar{l}_L & \bar{E}_L
\end{array} \right ) 
M \left(\begin{array}{c} l_R \\ E_R \end{array}
\right ) + h.c.
\qquad \quad
M = \left(
\begin{array}{c}
\end{array}
\begin{array}{cc}
M_D  & M_L\\
M_R & M_S 
\end{array} \right)\;,
\eeq
where the various entries are specified as
\beq
M_D = g_3 k' + g_4 k \, , \quad M_L = -g_7 v_L \, \quad  M_R = -g_7 v_R 
 \, , \quad  M_S =  g_8 \omega .
\eeq
Here the matrix $M_{D}$ is the Dirac mass term determined by the
standard Higgs bi-doublet VEV $\VEV{\phi}$ responsible for quark and
charged lepton masses, $M_L$ and $M_R$ are $G$ and $B-L$ violating
mass terms determined by $v_L$ and $v_R$, while $M_S$ is a gauge
singlet $G$-violating mass, proportional to the VEV of the gauge
singlet higgs scalar $\sigma$ carrying 2 units of $G$ charge.

In order to determine analytically the light charged-lepton masses and
majoron couplings we will work in the {\sl seesaw} approximation,
which we define as $M_S, M_R \gg M_D, M_L$. In this case the mass
matrix in \eq{MAT} can be brought to block diagonal form via a
bi-unitary transformation with two unitary matrices $V_L$ and $V_R$. In
the see-saw approximation, $V_L$ and $V_R$ can be written in the
form:
\beq
\label{ML}
V_L =
\left( \begin{array}{cc}
1-\frac{1}{2} \rho_L \rho_L^{\dagger} & - \rho_L \\
\rho_L^{\dagger} & 1- \frac{1}{2} \rho_L^{\dagger} \rho_L 
\end{array}
\right)\;,
\eeq
and
\beq
\label{MR}
V_R =
\left( \begin{array}{cc}
1-\frac{1}{2} \rho_R \rho_R^{\dagger} & - \rho_R \\
\rho_R^{\dagger} & 1- \frac{1}{2} \rho_R^{\dagger} \rho_R 
\end{array}
\right)\;,
\eeq
where $\rho_L$ is of order of the ratio of the lighter ($M_D$, $M_L$) to
the heavier scales ($M_R$, $M_S$), whereas
$\rho_R$ is of the order of the ratio $M_R/M_S$. Imposing that the
off-diagonal blocks vanish, the expression for $\rho_L$ is  
\beq
\label{rhoL}
\rho_L = (M_D M_R^{\dagger} + M_L M_S^{\dagger})
(M_R M_R^{\dagger} + M_S M_S^{\dagger})^{-1} + {\cal O}(\rho_L^3),
\eeq
Further simpification can arise if we assume either $M_S$ or $M_R$ to be
dominant. In the limit of $M_S \gg M_R$, which we will assume,
\beq
\label{rhoLapp}
\rho_L \approx M_L  M_S^{-1}+ M_D M_R^{\dagger}
{M_S^{\dagger}}^{-1} M_S^{-1},
\eeq
The expression for $\rho_R$ in the same limit is
\beq
\label{rhoR}
\rho_R \approx  M_R^{\dagger}
{ M_S^{\dagger}}^{-1}
\eeq

The block-diagonal form of the mass matrix resulting from the
bi-unitary transformation then is
\beq
M_{b.d.} = V_L M V_R^{\dagger} = \left(
\begin{array}{cc}
M_{Light}  &  0 \\
0  &  M_{Heavy}
\end{array}
\right) =
\left(
\begin{array}{cc}
M_D - M_L M_S^{-1} M_R  &  0 \\
0  &  M_S + \frac{1}{2} M_R M_R^{\dagger} {M_S^{\dagger}}^{ -1}
\end{array}
\right)
\label{massblock}
\eeq

Finally, each block of the matrix $M_{b.d.}$ is diagonalized by two
$3\times 3$ unitary matrices. Thus, for example, the light lepton
block $M_D - M_L M_S^{-1} M_R$ is diagonalized by the matrices $T_L$
and $T_R$:
\beq
T_L M_{Light} T_R^{\dagger} =
T_L ( M_D - M_L M_S^{-1} M_R ) T_R^{\dagger} = \mbox{diag}
(m_e,m_{\mu},m_{\tau})
\eeq

It can be seen that the light charged lepton masses are of the order
of the scale of $M_D$, i.e., the bi-doublet VEV's $k,k'$, with a small
correction $-M_L M_S^{-1} M_R$, whereas the heavy leptons get a mass
of the order of the scale of $M_S$, viz., the singlet VEV $\omega$.

\section{Flavour-violating majoron couplings to charged leptons}

The profile of the majoron in \eq{profilargew}  can be used, together
with the Yukawa couplings in \eq{yukawas} to write down the majoron
couplings to charged leptons in the mass-eigenstate basis,
defined as 
\beq
-{\cal L}_J = \frac{iJ}{\sqrt{2}} g_{ij}\bar{e}_{iL} e_{jR} + h.c.
\eeq 
The diagonalization procedure described in the previous section 
can then be used to determine the form of the $g_{ij}$ matrix elements
in the seesaw approximation.
We are particularly interested in the off-diagonal majoron 
couplings of the light charged leptons $e$, $\mu$ and $\tau$.
The result is:
\beq
-{\cal L}_J = \frac{iJ}{\sqrt{2}A} \bar{e}_L T_L \left[ V_L
\left( \begin{array}{cc}
\frac{v_L^2}{V^2 r} M_D & -\frac{v^2}{V^2 r} M_L   \\
   -\frac{v_L^2 v^2}{v_R^2 V^2 r} M_R & -M_S 
\end{array}
 \right) V_R^{\dagger} \right] _{11} T_R^{\dagger}  e_R + h.c. 
\label{majcoupl}
\eeq 
where
\beq
A = \left( \omega^2 + \frac{v_L^2 v^2}{V^2 r} \right)
^{\frac{1}{2} }
\eeq

On substituting for $V_L$ and $V_R$ from \eq{ML} and \eq{MR}, and
using the expressions for $\rho_L$ and $\rho_R$ from \eq{rhoLapp} and
\eq{rhoR}, respectively, we get
$$
-{\cal L}_J  \simeq 
\frac{iJ}{\sqrt{2}\omega} \bar{e}_L T_L [ \frac{v_L^2}{V^2 r}
 (M_D -M_L M_S^{-1} M_R )+ 
$$
\begin{equation}
+M_D M_R^{\dagger} {M_S^{\dagger}}^{-1} M_S^{-1} M_R
\left(1-\frac{v_L^2 v^2}{v_R^2 V^2} + \frac{v_L^2}{2 V^2 r}\right) ]
T_R^{\dagger} e_R + h.c.
\label{diagandoff}
\end{equation}
Since the first term in the square bracket is proportional to the 
light mass matrix itself, it gets diagonalized on multiplying by 
the matrices $T_L$ and $T_R^{\dagger}$. The second term in the 
square bracket gives rise to the first off-diagonal couplings
of the majoron to the charged leptons.

An alternative way to determine these couplings on more
general grounds is by using the Noether's theorem, i.e.
the method given in ref. \cite{774}.  The application
of this method of obtaining the majoron couplings to
the present situation is described in the Appendix.

Let us now proceed with the phenomenology of the off-diagonal
majoron coupling to charged leptons and estimate their 
magnitude within the seesaw approximation, in which there is
a hierarchy of scales, e.g. $\omega^2 \gg {v_R}^2$.

The dominant off-diagonal majoron coupling to a pair of charged leptons
$l_i l_j$ can be read from \eq{diagandoff}
\beq
g_{ij} \simeq \frac{1}{\omega}
\left(1-\frac{v_L^2 v^2}{v_R^2 V^2} + \frac{v_L^2}{2 V^2 r}\right)
[T_L M_D M_R^{\dagger} {M_S^{\dagger}}^{-1}
M_S^{-1} M_R T_R^{\dagger}]_{ij}
\eeq
Let us estimate the magnitude of this coupling for the case when one
of the leptons is $\tau$, and the other is either $e$ or $\mu$. For
this purpose we make a simplifying assumption, namely that assume that
the leading contribution to $g_{i\tau}$, where $i=e,\mu$, comes from
the diagonal entry of the product of $M_R$ and $M_S$ matrices,
$$
[M_R^{\dagger} {M_S^{\dagger}}^{-1}
M_S^{-1} M_R]_{lk} \simeq \delta_{lk} 
\left (\frac{v_R}{\omega}\right )^2
$$
The off-diagonal coupling is then
\beq
g_{i\tau} \simeq \frac{1}{\omega}
\left (\frac{v_R}{\omega}\right )^2
[T_L M_D T_R^{\dagger}]_{i\tau}
\eeq

Now since $M_D$ is the first term of the mass matrix of the light
charged leptons in \eq{massblock}, one can make the approximation
$[T_L M_D T_R^{\dagger}]_{i\tau} \approx m_\tau T_{Li\tau} T_{R\tau\tau}$.
Here the elements of the diagonalizing matrices $T_L$ and $T_R$ correspond
to the sin ($s_{i\tau}$) and cos ($c_{i\tau}$) of the $(i\tau)$
mixing angle. Therefore we can write
\beq
g_{i\tau} \simeq \frac{m_\tau}{\omega}
\left (\frac{v_R}{\omega}\right )^2
s_{i\tau} c_{i\tau}
\eeq
One can check that for reasonable values of the scales $v_R$ and
$\omega$, for instance $\omega \approx 10^6$ GeV and $v_R \approx 3
\cdot 10^5$ GeV, one gets this coupling as large as $10^{-8}$. Note 
that, as expected, the flavour violating majoron coupling is
suppressed for very small values of the mixing angle $(i\tau)$, since
it is proportional to $s_{i\tau}$. However, it is important to
realize that $g_{i\tau}$ can reach the relevant range where effects
are measurable $g_{i\tau} \sim 10^{-8}$ without requiring specially
large values of the mixing angle $(i\tau)$. Moreover, notice also that
this is consistent both with the VEV seesaw relation \eq{vevseesaw}
and with the astrophysical limit \eq{red} provided that $\lambda^2
\lsim 10^{-5}$ or, equivalently, provided that $v_L \lsim$ few
GeV. This brings the expected branching ratios (BR) for the decays
$\tau \to \mu J$ and $\tau \rightarrow e J$ in the range of
experimental observation in future experiments.

\FIGURE[t]{\psfig{file=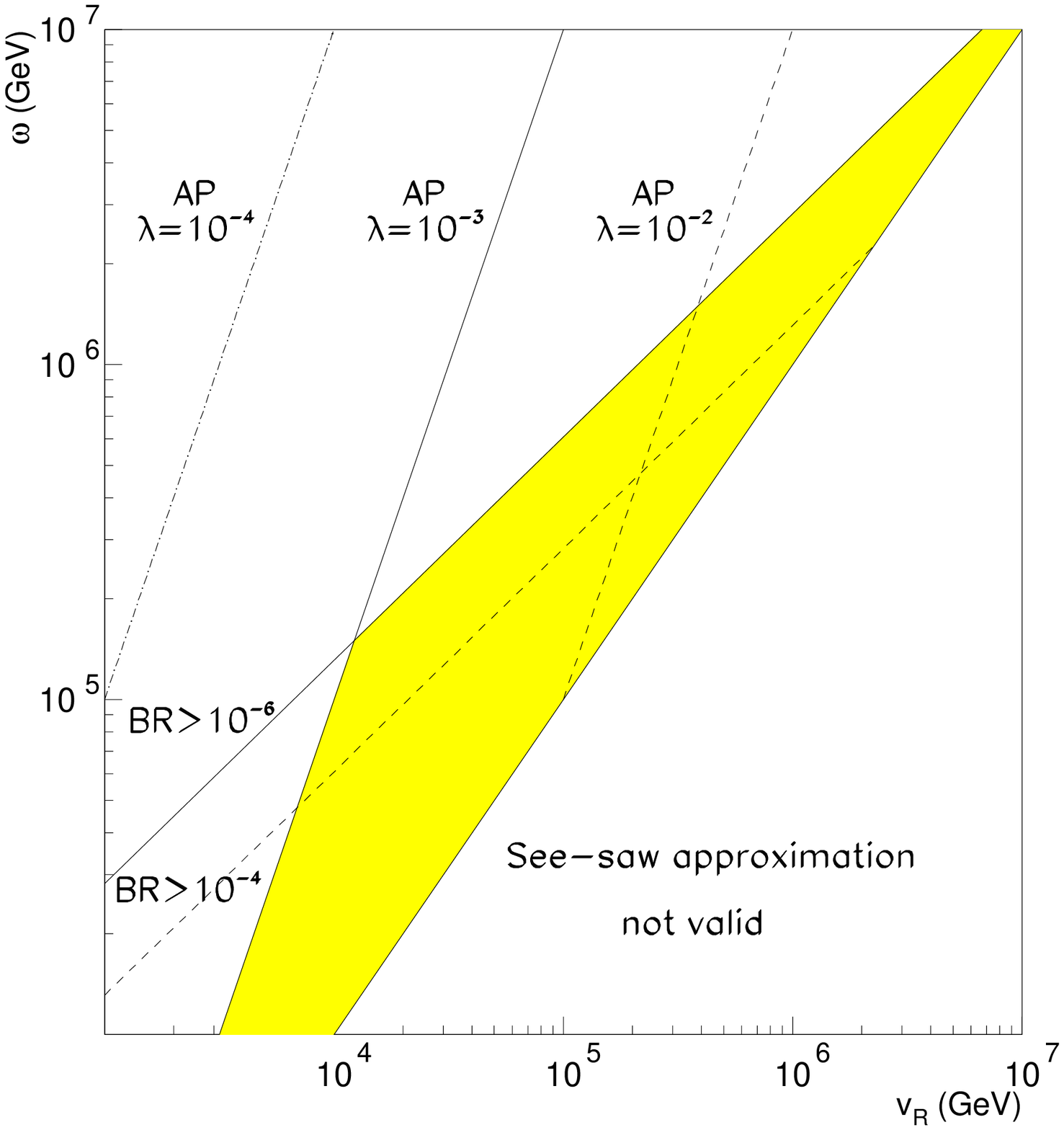, width=0.65\textwidth}
\caption{Region of parameters in the
$(v_R,\omega)$ plane that could lead to a majoron decay of the $\tau$
lepton with BR of the order $10^{-6}-10^{-4}$ in the see-saw
approximation.}
\label{ss}}
The above semi-analytic estimates are illustrated in figure
\ref{ss}. We display the region of relevant scale parameters 
$(v_R,\omega)$ that could lead to $\tau$ decays to majorons with BR of
the order $10^{-6}-10^{-4}$.  We assume that the product of the
relevant Yukawa couplings and diagonalizing matrix elements is of
order $10^{-1}$. The restrictions, shown as lines in the plot, come
from the astrophysical constraint of \eq{red0}, the requirement upon
the BR and the see-saw limit approximation (one should have at least
$\omega^2 \gsim v_R^2$). The condition from astrophysics requires
$v_R$ and $\omega$ to be on the right side of the lines labelled with
AP, shown for three different values of the parameter $\lambda$ in
\eq{vevseesaw}. On the other hand the BR of the majoron process can be
larger than $10^{-6}$ ($10^{-4}$) for values of the scales below the
thick solid (dashed) line. For definiteness, the shaded region
corresponds to BR larger than $10^{-6}$ for $\lambda \geq 10^{-3}$.

\subsection{Numerical Results}

\TABLE[l]{
\begin{tabular}{|c|c|}
\hline
Parameter & Range \\
\hline \hline
$\lambda$ & $10^{-4} \rightarrow 1$ \\
$\omega$ & $10^{4} \rightarrow 10^{8}$ GeV\\
$v_R$ & $10^{3} \rightarrow 10^{8}$ GeV\\
Yukawas & $10^{-5} \rightarrow 10^{-2}$\\
\hline
\end{tabular}
\caption{Ranges of variation of the different parameters used in our
numerical calculations.}
\label{table2}}
We now turn to a more accurate determination of the expected
branching ratios for the majoron emitting  $\tau$ and 
$ \mu$ decays, for the more general case in which the
hierarchy of scales used in the seesaw approximation is relaxed.
In order to do this we have performed the numerical 
calculation of the off-diagonal elements of the majoron 
couplings matrix in \eq{majcoupl} in the case that the
ratio $v_R/\omega$ is left arbitrary (no {\sl see-saw}). 
The various parameters such as $\lambda$, the symmetry 
breaking scales and Yukawa couplings were randomly varied
in certain reasonable ranges given in Table \ref{table2},
and the scales $v$ and $v_L$ are fixed by \eq{vevseesaw} and the
light $W$ boson mass.
For those choices which were not in conflict with the astrophysical
bound of \eq{red0}, we have calculated the corresponding branching
ratios for the lepton-flavour-violating decays of \eq{single}. Our
results for tau are summarized in \fig{fig1}
for the two possible channels and in  \fig{fig2} 
for the $\mu \to e + J$ case for four different ranges of $\lambda$
values. The points represent pairs of ($v_R,\omega$) values which lead
to the branching ratios larger than indicated. 

\FIGURE[t]{\psfig{file=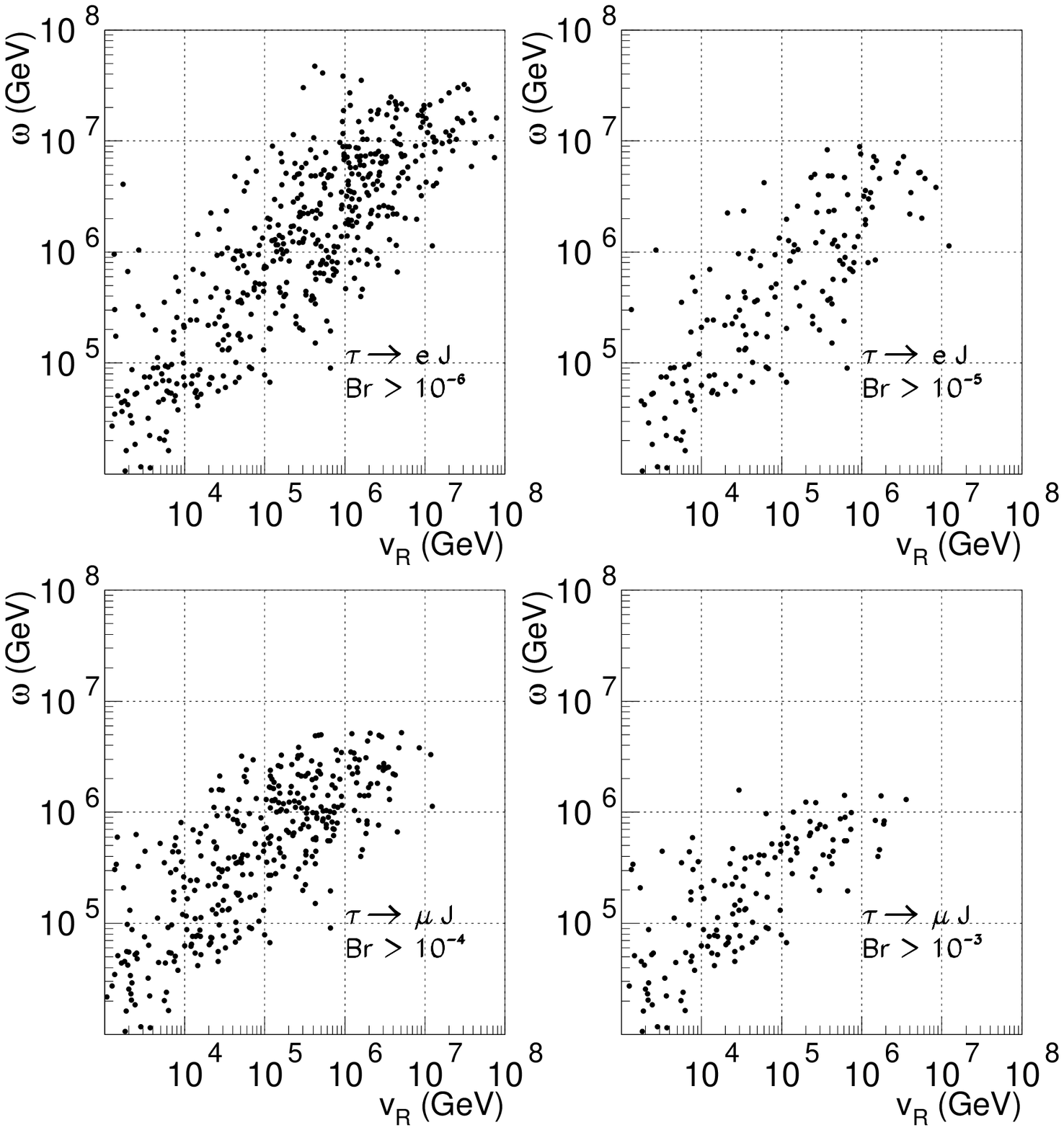, width=0.95\textwidth}
\caption{Allowed regions in the $v_R-\omega$ plane for $\tau$ decays.}
\label{fig1}}
\FIGURE[t]{\psfig{file=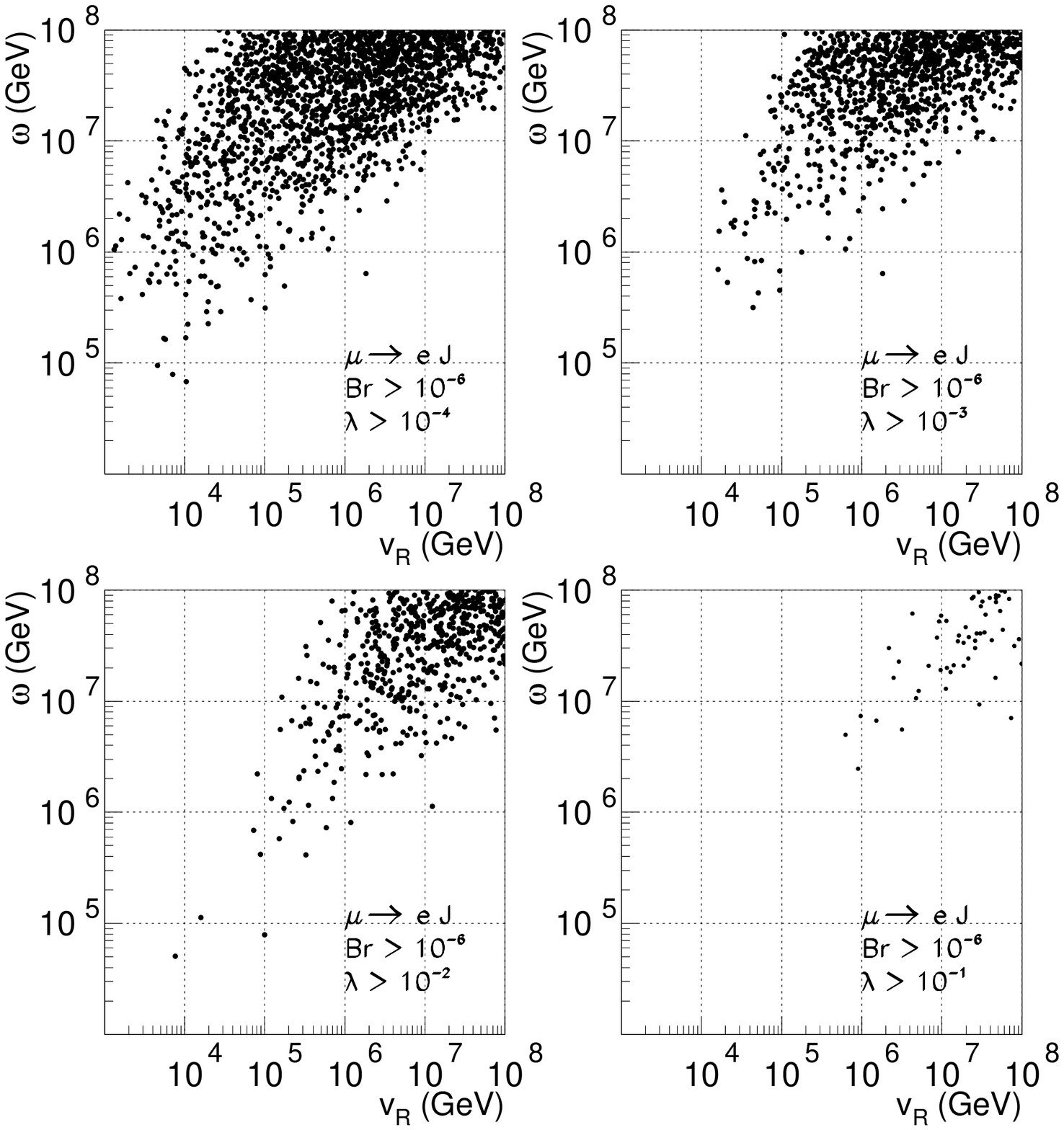,  width=0.95\textwidth}
\caption{Allowed regions in the $v_R-\omega$ plane for $\mu$ decays.}
\label{fig2}}
Clearly the $\tau \to e J$ branching ratios can reach or even exceed
$10^{-5}$. These values are in principle within reach of a possible
tau-charm factory which could obtain the limits $B(\tau \rightarrow
e J)<10^{-5}$ (standard optics) or $B(\tau \rightarrow e
J)<10^{-6}$ (monochromator \cite{monocrom}).  For the $\tau \to \mu
J$ case the branching ratios allowed in our model can reach or even
exceed $10^{-4}$. Though the attainable sensitivity in this case is
$B(\tau \rightarrow \mu J)<10^{-3}$, which is limited by $\mu/\pi$
separation, it may in principle be improved by an order of magnitude
with a RICH detector. For a discussion see ref. \cite{9307252} and
references therein. 

The allowed $\mu \to e J$ branching ratios are illustrated in
\fig{fig2} for various choices of the range of variation of
$\lambda$. The present experimental limit from TRIUMF \cite{mu} on
$\mu \to e + J$ is $BR(\mu \to e + J) \leq 2.6 \times 10^{-6}$.  One
sees that for the most favorable case studied, $\lambda \geq 10^{-4}$
we find that the $\mu \to e J$ branching ratios can reach or exceed
$10^{-6}$ which is quite close to the present limit. Presumably this
could be improved either at TRIUMF or PSI.

\section{Discussion and Conclusions}

We have proposed in this paper an extension of the $SU (2)_L \otimes
SU (2)_R \otimes U (1)_{B-L}$ Left-Right symmetric gauge model for the
electroweak interaction in which a global $U(1)_G$ lepton-number-like
symmetry is postulated.  Its spontaneous breaking at a scale $\omega
\gsim 10^4$ GeV or so would give rise to interesting physical effects
associated to the corresponding physical Nambu-Goldstone boson, called
majoron (denoted $J$).  Due to the existence of lepton-number-carrying
heavy isosinglet leptons $E_i$ there can be tree-level
flavour-violating majoron-charged-lepton couplings, leading to
sizeable majoron-emitting lepton-flavour-violating muon and tau
decays. We have discussed explicitly a simplest version of the model
and showed that the branching ratios for $\mu \to e+J$, $\tau \to e +
J$ and $\tau \to \mu + J$ decays can be large enough to fall within
the sensitivities of future $\mu$ and $\tau$ factories. On the other
hand the left-right gauge symmetry breaking scale may be as low as few
TeV. A simple variant of the model can be considered in which
isosinglet quarks instead of leptons are introduced, in which case one
would expect similar effects in the decays of hadrons, such as K
mesons.

Before concluding, note that the presence of the isosinglet charged
leptons $E_i$ also leads to flavour-violating couplings of the light
charged leptons to the neutral gauge bosons $Z_{1,2}$ present in the
model, already at the tree level. These would lead also to more
standard lepton-flavour-violating decays such as $\tau \to 3e$, $\tau
\to 3 \mu$, etc. Such decays are potentially interesting both for a
tau-charm factory as well as for a B meson factory
\cite{9307252}. However, we have investigated the corresponding
lepton-flavour-violating couplings to neutral gauge bosons in our model
and have found that the they are suppressed by factors of $v_L/\omega
\approx 10^{-5}$. This would lead to effects which are too small to
observe. Finally, photon emitting lepton-flavour-violating decays such
as $\mu \to e + \gamma$ in left-right symmetric models have been
widely discussed in the literature.  We avoid their discussion as it
would require the specification of the neutral lepton sector so that
the corresponding predictions would be more model-dependent. We
conclude that the search for lepton-flavour-violating majoron-emitting
weak decays provides a most sensitive way of testing our model.

\bigskip

\acknowledgments

This work was supported by the TMR network grant ERBFMRXCT960090 of
the European Union and by DGICYT under grant number PB95-1077.
S.D.R. was supported by the sabbatical grant SAB95-0175.

\bigskip

\appendix

\section{Appendix}

In this appendix we consider an alternative, more elegant 
and more general method to derive the off-diagonal couplings 
of the charged leptons to the majoron. This method is a 
two-fold generalization of the one given in ref. \cite{774}. 
Firstly, it is generalized to the case of Dirac fermions, 
and secondly, and  more importantly, it is generalized to 
the case where the majoron gets contributions from the 
neutral components of more than one scalar multiplet.

The first step is to write down the Noether current of the 
charged leptons in the weak basis and
the scalars corresponding to each conserved 
charge. Here we need consider only
the currents which are themselves electrically neutral. This gives
us the equations
\begin{eqnarray}
J_{\mu}^A =\sum^N_{i=1}\left(\bar{l}_{iL}\gamma _{\mu}l_{iL}g_{iL}^A
                        + \bar{l}_{iR} \gamma _{\mu} l_{iR} g_{iR}^A
                        + \bar{E}_{iL} \gamma _{\mu} E_{iL} G_{iL}^A
                        + \bar{E}_{iR} \gamma _{\mu} E_{iR} G_{iR}^A
					\right)
\nonumber\\
+ \sum_a \phi_a^R \stackrel{\leftrightarrow}
{\partial}_{\mu} \phi_a^I g_a^A
\end{eqnarray}
Here $A$ stands for each independent charge, the index $i$ runs over 
lepton flavours, and $a$ runs over the various neutral scalar fields. 
The superscripts R and I refer to the real and imaginary parts of 
the scalar fields, respectively.  $g_i^A$ and $G_i^A$ are the charges 
of the fields $l_i$ and $E_i$, with L and R 
denoting the chiralities, and $g_a^A$ are the charges of the 
scalar fields.

Conservation of the Noether currents corresponds to the equation
\begin{eqnarray}
\label{noether}
\partial ^{\mu}
\left[\sum^N_{i=1} \left( \bar{l}_{iL} \gamma _{\mu} l_{iL} g_{iL}^A
                        + \bar{l}_{iR} \gamma _{\mu} l_{iR} g_{iR}^A
                        + \bar{E}_{iL} \gamma _{\mu} E_{iL} G_{iL}^A
                        + \bar{E}_{iR} \gamma _{\mu} E_{iR} G_{iR}^A
					\right) \right]
\nonumber\\
+ \sum_a \langle \phi_a^R \rangle \Box \phi_a^I g_a^A=0
\end{eqnarray}
at the semiclassical level. In the case when the scalar field carry 
only a single  non-zero charge $g^A$, the combination 
$\langle \phi_a^R \rangle \phi_a^I g_a^A$ is the majoron field $J$,
 apart from a normalization factor.
The above equation then directly gives the coupling of the majoron to 
the various leptonic fields, using Lagrange's equation
\beq
\label{boxJ}
\Box    J = - \frac{\partial {\cal L}_J}{\partial J},
\eeq
and the Dirac equation on the left-hand side for the various
fermionic fields.
In the general case, the majoron is a linear combination 
\beq
J =  \sum_a \langle \phi_a^R \rangle \phi_a^I 
	(\sum_A \alpha_A g_a^A),
\eeq
In that case, multiplying 
\eq{noether} by $\alpha_A$ and summing over $A$ we get
\beq
\label{noetherold}
\partial ^{\mu}
 \left[ \sum_{i,A}\alpha_A \left( 
	\bar{l}_{iL} \gamma _{\mu} l_{iL} g_{iL}^A
    + \bar{l}_{iR} \gamma _{\mu} l_{iR} g_{iR}^A
    + \bar{E}_{iL} \gamma _{\mu} E_{iL} G_{iL}^A
    + \bar{E}_{iR} \gamma _{\mu} E_{iR} G_{iR}^A
	\right) \right] 
	+ \Box    J =0.
\eeq
Again using \eq{boxJ} the above equation gives the majoron couplings
in terms of the weak basis fields. To get the result for the
off-diagonal couplings, we first use the Dirac equation for the
fermionic fields to get
\beq
-i
 \sum_A\alpha_A 
( \! 
\begin{array}{cc} \bar{l}_L& \bar{E}_L \end{array}
 \! )\left\{
\left( \begin{array}{cc} g_L^A\, I & 0 \\
   0 & G_L^A\, I \end{array} \right)M - M 
\left( \begin{array}{cc} g_R^A \, I & 0 \\
   0 & G_R^A \, I \end{array} \right)\right\}  
\left( \begin{array}{c} l_R\\ E_R \end{array}
\right) + h.c.
	- \frac{\partial {\cal L}}{\partial J} =0.
\eeq
On expressing the weak basis fields in terms of the diagonal fields,
this would give the complete expression for the leptonic majoron
couplings.  However, if we are interested only in the off-diagonal
couplings, we can achieve a simplification by noting that if the
quantum numbers $g$ and $G$ in the above equation were equal, the
couplings of the majoron would be just proportional to the mass
matrices, and therefore diagonal. Hence the deviation from diagonality
of the couplings would be proportional to the differences between $g$
and $G$. Using this fact, the equation for the off-diagonal couplings
then reduces to:
\beq
{\cal L}^J_{off-diag} = -i J
 \sum_A\alpha_A \\ 
(\!  
\begin{array}{cc} \bar{l}_L& \bar{E}_L \end{array}
\! )\left\{
\left( \begin{array}{cc} 0 & 0 \\
   0 & \delta g_L^A\, I \end{array} \right)M - M 
\left( \begin{array}{cc} 0 & 0 \\
   0 & \delta g_R^A \, I \end{array} \right) 	\right\} 
\left( \begin{array}{c} l_R\\ E_R \end{array}
\right)
	+ h.c.,
\eeq
where 
\beq
\delta g_L^A = G_L^A - g_L^A ; \quad 
\delta g_R^A = G_R^A - g_R^A. 
\eeq
Using the relations between the weak and mass basis states for the
light leptons we then get the off-diagonal majoron couplings as
$$
{\cal L}^J_{off-diag} = -i J
 \sum_A\alpha_A 
\bar{e}_{iL}
(T_L)_{ij}\{\left[(V_L)_{12}(V_L)^{\dagger}_{12}\right]_{jk}
m_k (G_L^A-g_L^A) -
$$
\beq
-m_i (G_R^A-g_R^A) \left[(V_R)_{21} (V_R)^{\dagger}_{21}\right]_{jk}
\} (T_R)^{\dagger}_{kl} e_{lR} + h.c.
\label{offdiag}
\eeq
Knowing the diagonalizing matrices $T_L$, $T_R$, $V_L$ and $V_R$ in a 
specific model, the
off-diagonal couplings can easily be calculated.
 
We illustrate this method for the case at hand. In our specific
model,  the various independent conserved charges can be chosen as
the global $U(1)$ charge, $B-L$ and the combination $T_{3L}-T_{3R}$. 
The quantum numbers of the various fields are given in Table \ref{table1}. 
The majoron is the linear combination of the neutral scalar fields
given in \eq{profi}.  This implies the values of $\alpha_A$
as follows:
\beq
\alpha _1 = \frac{1}{N},\; \; 
\alpha_2 = \frac{v^2 (v_L^2 - v_R^2)}{V^2v_R^2rN}, \; \;
\alpha_3 =  \frac{-2 v_L^2}{ V^2 rN},
\eeq

where 
\beq
N   = 2 ( \omega^2 + \frac{v_L^2 v_R^2 v^2}{V^2 v_R^2 + v^2 v_L^2} )
^{-\frac{1}{2}}.
\eeq
Substituting these values and the various quantum numbers in 
\eq{offdiag} gives
$$
{\cal L}^J_{off-diag} = -i J
 \bar{e}_{iR} \{ m_k \left( T_L (V_L)_{12}(V_L)_{12}^{\dagger}
T_R^{\dagger} \right)_{ik} \left(\frac{-2 v_L^2 (v^2+v_R^2)}{v_R^2 V^2 r}
 \right)
$$
\beq
 - m_i \left(
 T_L (V_R)_{12}(V_R)_{12}^{\dagger}
T_R^{\dagger} \right)_{ik} \left(\frac{2}{r}
 \right) \} e_{kL} + h.c. 
\eeq
The final result given in \eq{diagandoff} can then be obtained
on substituting for $V_L$ and $V_R$ from \eq{ML} and \eq{MR}.



\begin{thebibliography}{99}

\bibitem{LR1}
J.C. Pati and A. Salam, \prd{10}{1975}{275}; 
R.N. Mohapatra and J.C. Pati, \prd{11}{1975}{566}, \prd{11}{1975}{2558}. 

\bibitem{LR2}
R.N. Mohapatra and G. Senjanovi\'c, \prd{23}{1981}{165}. 

\bibitem{RHOCRIT}
S. Gerstein and Ya.B. Zeldovich, {\it Zh. Eksp. Teor. Fiz. Pisma. Red. }{\bf 4} (1966) 174;
R. Cowsik and J. McClelland, \prl {29}{1972}{669};
D. Dicus et al., \apj {221}{1978}{327};
B.W. Lee and S. Weinberg, \prl{39}{1977}{165}.

\bibitem{CMP}
Y. Chikashige, R.N. Mohapatra and R. Peccei, \prl{45}{1980}{1926}; 
\plb{98}{1981}{265}.

\bibitem{V} 
J.W.F. Valle, \plb{131}{1983}{87};
G. Gelmini and J.W.F. Valle, \plb{142}{1984}{181};
J.W.F. Valle, \plb{159}{1985}{49};
M.C. Gonz\'alez-Garc\'{\i}a and J.W.F. Valle, \plb{b216}{1989}{360};
A. Joshipura and S. Rindani, \prd{46}{1992}{3000}.

\bibitem{CON}
M.C. Gonz\'alez-Garc\'{\i}a and J.W.F. Valle, \plb{216}{1989}{360}.

\bibitem{fae}
J.W.F. Valle, {\it Gauge Theories and the Physics of
Neutrino Mass}, \ppnp{26}{1991}{91-171} and references therein.

\bibitem{DPRV}
A.D. Dolgov, S. Pastor, J.C. Rom\~ao and J.W.F. Valle, 
\npb{496}{1997}{24}.

\bibitem{bbnothers}
M. Kawasaki et al., \npb{419}{1994}{105};
S. Dodelson, G. Gyuk and M.S. Turner, \prd{49}{1994}{5068};
S. Hannestad, \prd{57}{1998}{2213};
M. Kawasaki et al. , K. Kohri and K. Sato, \plb{430}{1998}{132}.

\bibitem{Raffelt0}
G. Raffelt, {\it Stars as Laboratories for Fundamental
Physics} (The University of Chicago Press, 1996).

\bibitem{previous}
E.Kh. Akhmedov, A.S. Joshipura, S. Ranfone and J.W.F. Valle,
\npb{441}{1995}{61}; for an early alternative proposal in this
direction see, A. Kumar and R.N. Mohapatra, \plb{150}{1985}{191}.

\bibitem{lqsst} 
S. Kalara and K.A. Olive, \npb{331}{1990}{181}; 
J. Distler, B.R. Greene, K.H. Kirklin, and P.J. Miron, \cmp{122}{1989}{117}.

\bibitem{lqphen}
V. Barger, N.G. Deshpande, and K. Hagiwara, \prd{36}{1987}{3541};
\ijmpa{2}{1987}{1181}.

\bibitem{lqphen2}
V. Barger et al in Proceedings, {\sl Physics of the Superconducting
Supercollider}, Snowmass 1986, p. 216-220.

\bibitem{univseesaw}
A. Davidson and K. C. Wali, \prl{59}{1987}{393}. For a recent
paper see, e.g.  Yoshio Koide, \prd{57}{1998}{5836} and
references therein.

\bibitem{NPBTAU}
J. C. Rom\~ao, N. Rius and J. W. F. Valle, \npb{363}{1991}{369}.

\bibitem{mu}
A. Jodidio et al., \prd{34}{1986}{1967}.

\bibitem{tau}
MARK III Collaboration, \prl{55}{1985}{1842}.

\bibitem{774}
J. Schechter and J.W.F. Valle,  \prd{25}{1982}{774}. 

\bibitem{CLEO97}
For a recent experimental discussion of tau decays by the CLEO
collaboration, see \prd{57}{1998}{5903}.

\bibitem{9307252}
For a discussion of the experimental prospects for detecting
lepton-flavour-violating muon and tau decays at future B and tau-charm
factories see {\sl Searching for Exotic Tau Decays}, by R. Alemany,
J.J. G\'omez-Cadenas, M.C. Gonz\'alez-Garc\'{\i}a and J.W.F. Valle,
preprint \hepph{9307252} or CERN-PPE/93-49 and references therein.

\bibitem{SST1}
R.N. Mohapatra and J.W.F. Valle, \prd{34}{1986}{1642};
I. Antoniadis et al., \plb{208}{1988}{209};
E. Papageorgiu and S. Ranfone, \plb{282}{1992}{89}.

\bibitem{WYLER}
D. Wyler and L. Wolfenstein, \npb{218}{1983}{205}.

\bibitem{DEAR}
For a review see J.E. Kim, \prep{150}{1987}{1}.

\bibitem{monocrom}
See for example, A. Faus-Golfe and J. Le Duff, Report LAL/RT 92-01 (1992).

\end{thebibliography}
\end{document}